\documentclass[aip,reprint,twocolumn,groupedaddress]{revtex4-1}
\usepackage{color}
\usepackage{graphicx}
\usepackage{amsmath}
\usepackage{latexsym}
\usepackage{graphicx}
\usepackage{grffile}
\usepackage{amssymb}
\usepackage{amsfonts}
\usepackage[autostyle]{csquotes}
\usepackage{bm}

\begin{document}
\title{Complex behavior in chains of nonlinear oscillators}
\author{Leandro M. Alonso}
\affiliation{New York, NY 10028, USA.}\altaffiliation{Current address: Volen Center for Complex Systems, Department of Biology, Brandeis University, Waltham, MA 02454, USA.}
\email{leandro.alonso.ruiz@gmail.com, lalonso@brandeis.edu}

\date{\today}
\begin{abstract}
This article outlines sufficient conditions under which a one-dimensional chain of identical nonlinear oscillators can display complex spatio-temporal behavior. The units are described by phase equations and consist of excitable oscillators. The interactions are local and the network is poised to a critical state by balancing excitation and inhibition locally. The results presented here suggest that in networks composed of many oscillatory units with local interactions, excitability together with balanced interactions are sufficient to give rise to complex emergent features. For values of the parameters where complex behavior occurs, the system also displays a high-dimensional bifurcation where an exponentially large number of equilibria are borne in pairs out of multiple saddle-node bifurcations. 
\end{abstract}
\maketitle

\begin{quotation}
What are the conditions under which large assemblies of oscillatory units will display complex behavior? This article explores minimal assumptions under which such a scenario is attainable in a network of identical nonlinear oscillators with local interactions. For open ranges of the parameter values, the network can display patterns of activity that spread from traveling pulses to complex emergent structures reminiscent of those that arise in cellular automata. The complexity of the patterns is quantified in a range of parameter space revealing transitions among at least three phases: an absorbing phase in which all initial conditions decay to equilibrium, a chaotic phase in which all initial conditions quickly evolve into disordered states, and a third phase in which the network displays complex behavior. The transition between the absorbing and complex phases can be partly explained by the presence of a high-dimensional bifurcation where an exponentially large number of equilibria are created in multiple saddle-node bifurcations. The model presented here can be a useful case study to explore the dynamical mechanisms by which complex behavior arises in spatial arrangements of oscillatory units with local interactions. 
\end{quotation}

\section{Introduction}
Rhythmic phenomena are ubiquitous across nature. Oscillations in physical, biological, and chemical systems can be expressed mathematically as periodic solutions of nonlinear dynamical systems $\dot{x} = f(x)$ \cite{gucken}. When autonomous oscillators interact with an external signal they adjust their amplitudes and phases. Under weak coupling assumptions the amplitudes remain relatively constant and the oscillators can be described by phase equations \cite{synchronization}. Phase oscillators are widely utilized to study interactions in assemblies of oscillators because they capture the dynamical features of a complicated nonlinear oscillator while being more amenable to analytic inquiry. Reducing the description of a general oscillator to a phase equation is useful in understanding the effect of interactions in large sets of oscillatory units, as most notably illustrated by the Kuramoto model \cite{kuramoto,strogatz, acebron}. The study of interactions in networks of oscillators has recently received increased attention due to a breakthrough by Ott et. al that allows us to obtain analytic results on the average properties of large sets of globally coupled phase oscillators \cite{ott,alonso11,roulet16}. However, less is known about the case in which the interactions are local and this is consequently an area of active research \cite{wolfrum,kopell}. 

It is generally accepted that periodic autonomous processes underlie the coordination of biological phenomena. Thus, a longstanding question in theoretical biology has been how such coordinated behavior emerges from the aggregate activity of many oscillatory units \cite{winfree}. In several cases, the dynamical features of these units in isolation are well-described, however, understanding the mechanisms by which complexity springs from \emph{interacting} sets of such units constitutes a very difficult problem. It has been argued that the origins of such complexity can be addressed by mathematical models known as \emph{Cellular Automata} (CA) \cite{wolfram}. These models consist of a large number of units that can take a finite number of states and evolve according to simple rules. Despite their discrete nature, CAs have been extensively utilized to model biological phenomena because of their computational properties (for a review see Ermentrout et. al \cite{ermentroutca}). In this article I introduce a time-continuous model of coupled oscillators whose spatio-temporal dynamics share some features with those of CAs and may therefore provide insight into the dynamical mechanisms by which complexity arises in systems composed of many oscillatory units. 

Recent developments in the experimental investigation of biological networks support the 
notion that many biological systems operate at or near critical regimes \cite{bialek}. These regimes are characterized by power law distributions of observable quantities, long-range correlations, and are typically associated with thermodynamic phase transitions. An example was recently derived from quantitative measurements of flocks of birds. While flocks travel coherently with a well-defined mean velocity, fluctuations over this mean level are correlated over long distances. The statistical properties of the flock can be explained quantitatively by minimally structured models whose parameters can be inferred from data, and are found to be in a critical regime \cite{bialekflocks}. Despite the interactions being local, long-range correlations in the critical regime allow for the propagation of information across distant sites in the flock. Recently, a connection between statistical criticality and dynamical criticality was introduced by Magnasco et. al \cite{magnascoprl}. They considered an abstract model of neurons with dynamic interactions which self-organizes towards a state that exhibits various long-tailed statistical observables. In their model, the activity of the units is encoded in a vector $\boldsymbol{x} \in R^N$ that evolves according to a linear equation $\boldsymbol{\dot{x}}=\boldsymbol{W}\boldsymbol{x}$, while the connectivity matrix $\boldsymbol{W} \in R^{N \times N}$ also evolves at a slower pace following an anti-hebbian rule $\boldsymbol{\dot{W}} = \tau (\boldsymbol{I_d} - \boldsymbol{x} \boldsymbol{x^t})$: if the activities of $x_i$ and $x_j$ are correlated, their connectivity decreases. The effect of this rule is that the anti-symmetric part of the connectivity matrix is conserved and the symmetric part evolves in such a way that the real part of all the eigenvalues $\lambda$ of $\boldsymbol{W}$ oscillate close to the critical value $Re(\lambda)=0$. Their model links the presence of a large number of dynamical modes of activity with marginal stability to long-range correlations and statistically critical behavior, and introduces an explicit mechanism by which such critical states can be achieved and supported.

The purpose of this article is to explore sufficient conditions under which an assembly of oscillatory units can display complex behavior. For this I present a simple model of coupled phase oscillators that --- upon changes of its parameters --- can display a multitude of patterns including traveling waves, spatio-temporal chaos, and most surprisingly, long-lived complex structures reminiscent of those that occur in class IV CAs. In terms of the complexity of its solutions, the model displays at least three well-differentiated regimes with marked transitions in parameter space. An analysis of existence and stability of equilibria shows that, upon changes of its parameters, the system undergoes a high-dimensional bifurcation in which an exponentially large number of equilibria are created locally. 

This work is organized as follows. The model is introduced in Section 2. The solutions of the model are explored numerically and their complexity is quantified in Section 3A. Section 3B contains an analysis of existence and stability of equilibria. In Section 3D, the results are extended to lattices of higher dimensions and the case of a two-dimensional lattice is presented as a case study for pattern formation. Finally, Section 4 contains the conclusions and future directions.

\section{The model}
The model explored in this article consists of a network of excitable units arranged in space along a one-dimensional lattice. Such arrangement is also known as a chain, or ring in the case of periodic boundary conditions. Chains of coupled phase oscillators have been studied in several contexts including computational models for motor pattern generation \cite{holmes}. Similar arrangements have been explored for other types of oscillators (see ref. \cite{pikovski} for relaxation oscillators). The model presented here was conceived in an attempt to explore the interplay between two properties: that the units are excitable and that long-range correlations are possible despite interactions being local.

In this model, the dynamics of the units is given by the Adler's equation \cite{adler}:  the state of a unit is determined by its phase $\theta_A$ and evolves according to $\dot{\theta}_A = \omega_A + \gamma_A \cos(\theta_A)$. This equation is widely utilized to model excitable oscillators and it captures the main dynamical features of an oscillator close to a Saddle-Node in Limit Cycle bifurcation (SNILC). Alternatively, it can be thought of as describing the phase difference of two oscillators coupled by their phase difference. The dynamics of the units in the absence of interactions is represented in phase space in Figure (1A). For values of $\omega_A<\gamma_A$ a unit displays a pair of equilibria that stem from a saddle-node bifurcation. For values of $\omega_A>\gamma_A$ the units display oscillations and if $\omega_A<\gamma_A$ all initial conditions decay to the stable state.   In the stable state, small perturbations lead to small responses, but if the perturbation is large enough the response of the unit corresponds to a large excursion in phase space. 

One mechanism to enable long-range correlations in the network it is to have a large number of dynamical modes with marginal stability \cite{magnascoprl}. As a way to explore the consequences of this mechanism, Magnasco et. al proposed an abstract model of cortical dynamics, in which they assumed that the interactions between units are local and that inhibition and excitation are exactly balanced in every local patch. This condition poises their network in what the authors term a massive high-dimensional Hopf bifurcation: the dynamics of their system is given by $\boldsymbol{\dot{x}} = \boldsymbol{A} \boldsymbol{x} + \boldsymbol{x^3}$ where $\boldsymbol{x} \in R^N$ is a vector, $\boldsymbol{x^3}$ is a local nonlinearity (in component notation $\boldsymbol{x^3}=x_{ij}^3$) and $\boldsymbol{A} \in R^{N\times N}$ is an antisymmetric matrix that couples the activity of neighboring units. The dynamics of their system to first order is given by $\boldsymbol{A}$ and therefore all modes of activity are marginal ($Re(\lambda) =0$). Under these assumptions their system displays interesting properties such as input-strength dependent spatial integration \cite{magnascoyan}. The model presented in this article explores the possibility that the same strategy of balancing the strength of interactions locally can be applied to enable long-range correlations in a network of excitable units. 

The assumptions in the model presented here can be summarized as follows: the units are excitable phase oscillators placed along a one-dimensional lattice, they interact only with their nearest-neighbors and excitation and inhibition are balanced in local patches. The state of the system can be specified by a $N$-dimensional vector $\boldsymbol{\theta}$ with components $\theta_i$ ($i \in [1,N]$) and the equations of motion are,
\begin{eqnarray}
  \dot{\theta_i}&=& \omega + \gamma \cos(\theta_i) + k (-1)^i \{\cos(\theta_{i-1}) + \cos(\theta_{i+1})\}. 
\label{model}
\end{eqnarray}
Parameter $\omega$ controls the natural frequency of the oscillators in the absence of the excitation term. The excitation term is controlled by parameter $\gamma$ and it can be regarded mathematically as a first order Fourier expansion of a more general vector field in the circle. The interaction between units is assumed to depend on the absolute value of the phases. The strength of the interactions is given by $k$ and the term $(-1)^i$ implements the assumption of balancing excitation and inhibition in local patches. This is achieved by arranging the units in an alternating fashion according to the sign of their interaction strength, along a one-dimensional axis, as shown in Fig. \ref{model}B. With this definition, any given unit is either purely excitatory or purely inhibitory. In order to simplify the analysis, we assume periodic boundary conditions by setting $\theta_1 = \theta_N$. Finally, we assume that $N$ is even so that interactions are balanced everywhere.

The solutions of system (\ref{model}) are discussed in the next section. The solutions of the model are first explored numerically in Section 3A: the complexity of the solutions is quantified in a range of parameter space. Section 3B contains an analysis of existence and stability of equilibria that links complex behavior to a high-dimensional bifurcation. For clarity, the analysis of existence and stability is discussed for the case of a one-dimensional chain. However, the model and the analysis can be extended to lattices of higher dimension. In section 3D, system (\ref{model}) is extended to two-dimensional lattices leading to an interesting case study of pattern formation.  

\begin{figure}[ht!]
\includegraphics[width=86mm]{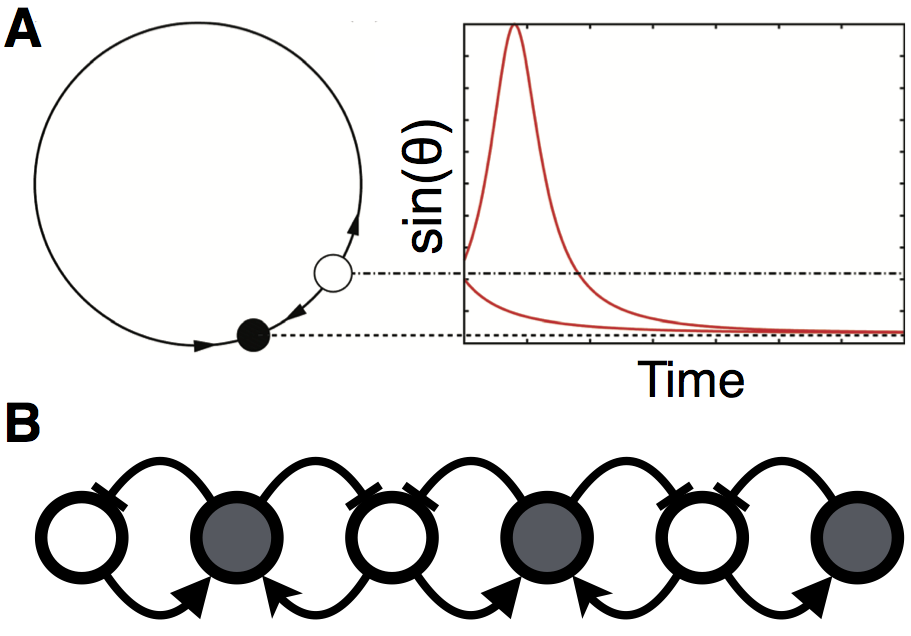}
\caption{\textbf{The model.} The model consists of a network of excitable oscillators placed along a one-dimensional lattice. The units can interact with their nearest-neighbors only, and the strength of the interactions is balanced everywhere. \textbf{A} The dynamics of the units in absence of interactions is given by the Adler equation: $\dot{\theta}_A = \omega_A + \gamma_A \cos(\theta_A)$. When $\gamma_A > \omega_A$ the units display a pair of stable and unstable equilibria. For initial conditions below the threshold (indicated by the dashed line), the system decays to the stable equilibria and for initial conditions above this threshold the system performs a large excursion in phase space. Thus, small perturbations of the stable state result in quiescence and large perturbations yield large responses. \textbf{B} The network is composed of $N$ identical units with nearest neighbor interactions. The units are represented by circles and the lines represent the interactions. The arrows correspond to excitatory (positive) coupling and \textbf{T}s correspond to inhibitory (negative) coupling. The shading indicates that each unit is either purely excitatory or purely inhibitory. Excitation and inhibition are balanced locally by placing alternating excitatory and inhibitory units along the one-dimensional lattice.}
\end{figure}  

\section{Results}
The solutions of system (\ref{model}) were explored numerically in the range $(\omega,\gamma,k=1) \in [0,L]$ with $L=3$. The choice of $L=3$ is useful to explore the case $\omega = \gamma$. While for the cases $\omega>0$, $\gamma>0$ and $k>0$ a parameter can be eliminated by rescaling time, it will be useful to include the limit cases given by $\omega=0$, $\gamma=0$ and $k=0$, since these parameters have intuitive physical interpretations. The simulations were done using a standard Runge-Kutta O(4) routine with fixed step $dt=0.01$ and the \emph{activity} is defined as 
$x_{i}=\sin(\theta_i)$. 

The results in this article are organized as follows. In Subsection 3A it is established that: (a) system (\ref{model}) can produce a multiplicity of qualitatively different spatio-temporal patterns, (b) that quantification of the complexity of such patterns can be used to construct a phase diagram in parameter space, and (c) that there are at least three different phases. In Subsection 3B, an analysis of existence and stability of equilibria is performed to demonstrate that system (\ref{model}) features a high-dimensional bifurcation in which, upon changes of the parameters, an exponentially large number of equilibria are created locally in pairs at multiple saddle-node bifurcations. The existence and stability of equilibria is compared to the phase diagram suggesting a plausible explanation for the main observed transition. Finally, Subsection 3D shows that system (\ref{model}) can be extended to two-dimensional lattices with similar results.

\subsection{Phenomenology}
\begin{figure*}[ht!]
\includegraphics[width=180mm]{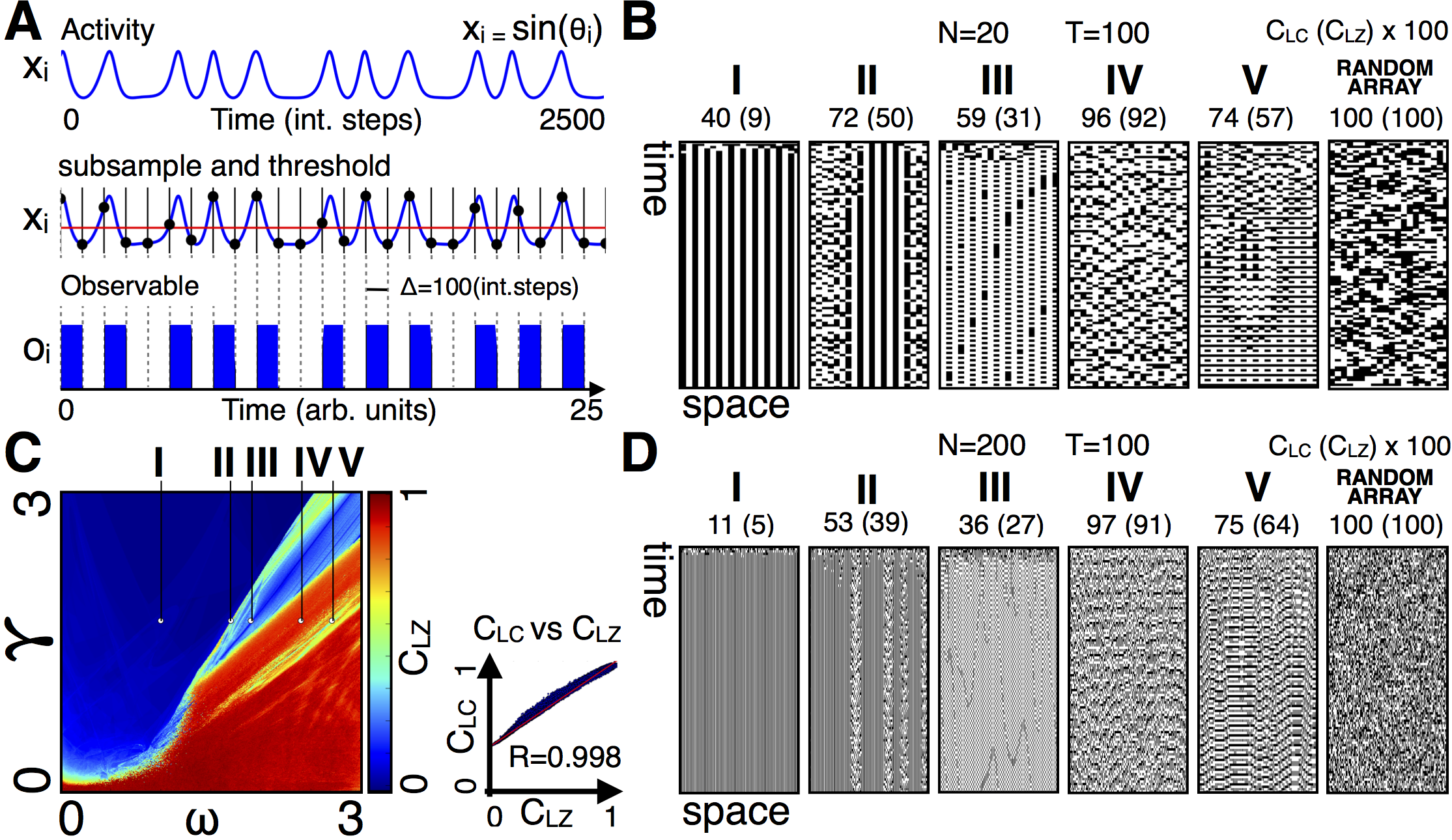}
\caption{\textbf{Methods.} The solutions of system (\ref{model}) are collapsed into a binary array $O$, and the complexity of $O$ is measured by standard procedures. \textbf{A} (top row) Activity of a representative unit. The activity of each unit is defined as $x_{i}=\sin(\theta_i)$. (center row) The activity is subsampled by a factor $100$ and collapsed to binary values using the mean activity as a threshold $<x_{i,j}>$ (red line). (bottom row) Representative column of the resulting boolean array $O$. \textbf{Β} \textbf{B} Observables $O$ for $N=20$ and $T=100$ for the values indicated in \textbf{C}: in all panels, $(\gamma=1.7)$, I $\omega=1$, II $\omega=1.7$, III $\omega=1.9$, IV $\omega=2.4$,V $\omega=2.7$ and a random boolean array for comparison (see main text). The labels indicate the values of $C_{LC}$ ($C_{LZ}$) for the patterns shown. \textbf{C} Lempel-Ziv complexity $C_{LZ}$ of $O$ for one random initial condition as a function of parameters using $N=20$ units and $T=100$ time samples. The dots and roman labels indicate points in parameter space with different values of $C_{LZ}$. The Lempel-Ziv complexity is compared against the \emph{lossless compression} complexity $C_{LC}$ yielding a linear relationship $C_{LC} \approx 0.67 C_{LZ} + 0.34$ with regression coefficient $r=0.998$. \textbf{D} Observables $O$ for $N=200$ and $T=100$ for the same parameter values as in \textbf{C}.}
\label{methods}
\end{figure*}  

System (\ref{model}) possesses many dynamical attractors and, in some regions of parameter space,  different initial conditions lead to very different spatio-temporal patterns: some initial states will quickly decay to stable equilibria or periodic attractors, while other initial states can trigger a much richer spatio-temporal evolution due to coexisting high-dimensional attractors. Here we derive a complexity measure that is able to identify parameter values for which the evolution of random initial states lead to qualitatively different spatio-temporal patterns. For the purposes of this article, it is sufficient to differentiate at least three different kinds of spatio-temporal patterns which will be termed as follows: absorbing, for quickly decaying solutions, complex, for solutions that trigger long complicated transients which may or may not decay to periodic attractors, and chaotic, for solutions that quickly evolve into disordered states with no evident regularities. In order to quantify the degree of regularity in a given solution, we adopt a compression-based approach: intuitively, an observable derived from an absorbing solution has a much simpler description than a chaotic pattern, and it should therefore be more compressible in an information theoretic sense. Compression-based approaches are commonplace in the analysis of complex systems and a formal discussion can be found in ref. \cite{vitanyi}. 

Figure \ref{methods} summarizes the methods used in Subsection 3A. In order to measure the complexity of a solution we take two steps: the solution is collapsed to a binary array $O$ and then the complexity of $O$ is measured by standard procedures. To construct $O$ from an initial state, the system is evolved for $T$ units of time and its \emph{activity} is stored in an $N \times \frac{T}{dt}$ array $X$. Array $X$ is subsampled by a factor $100$ and collapsed to binary values using the mean value of the activity $<X_{i,j}>$ as a threshold, yielding an observable binary array $O$ with $N \times \frac{T}{100 dt}$ values. This operation is illustrated in Figure \ref{methods}A. To provide a visual reference, Figure \ref{methods}B shows the observables $O$ of a network of size $N=20$ using $T=100$ samples for different parameter values. 

The complexity of the resulting binary arrays $O$ was measured by two procedures, both based on the Lempel-Ziv algorithm for lossless data compression \cite{lempelziv}. The first procedure consists of computing the Lempel-Ziv (LZ) complexity: a calculable measure of algorithmic complexity related to the LZ algorithm that is useful for characterizing spatiotemporal patterns in high-dimensional nonlinear systems \cite{kaspar}, and also for distinguishing among different types of cellular automata (a recent study using this measure on CAs can be found in ref. \cite{estevez-rams}). Closely related to this quantity, the second procedure used in this article consists of measuring the performance of a universal compressor at compressing $O$ \cite{zenil}. Let $X$ be a boolean array, $LZ(X)$ its LZ complexity, and $LC(X)$ its size in bytes after \emph{lossless compression} LC. Let $R$ be a random boolean array of the same shape as $X$, we define the complexities, 
\begin{eqnarray}
C_{LZ}(X) = \frac{LZ(X)}{<LZ(R)>}\\ \nonumber
C_{LC}(X) = \frac{LC(X)}{<LC(R)>},
\end{eqnarray}
where the normalization constants $<LZ(R)>$ and $<LC(R)>$ correspond to the average over $100$ realizations of the random array $R$. Figure \ref{methods}C shows the values of $C_{LZ}$ in parameter space for a network of size $N=20$ computed over a regularly spaced $501 \times 501$ grid in the inspected range. The observables $O$ were constructed by evolving the \emph{same} initial condition (taken at random) for $T=100$ units of time. The values of the parameters used to build the observables shown in Fig. \ref{methods}B are indicated by white dots and they were chosen in regions where $C_{LZ}$ takes different values to provide a reference. With these choices, $C_{LZ}$ is sufficient to distinguish a regular pattern such as (I) from a chaotic pattern such as (IV), and also to identify patterns that present more regularities like (III) and (V). 

The Lempel-Ziv complexity LZ can be calculated with a short algorithm \cite{kaspar}. However, a major drawback for the purposes of this article is that the time it takes to compute LZ for a random string of length $n$ scales as $n^2$, posing a difficulty for the numerical exploration of larger networks \cite{kaspar}. In order to compare LZ against LC, a least squares fit was performed returning $C_{LC} \approx 0.67 C_{LZ} + 0.34$ with regression coefficient $r=0.998$ (see Fig. \ref{methods}C). The two measures are in agreement, however, $C_{LC}$ is computationally more efficient, it can be reproduced using standard compressors and offers an intuitive interpretation. In this article LC was implemented using the compressor of the popular python library numpy and it is reasonable that $C_{LC} \propto C_{LZ}$ since most compression algorithms are based on the Lempel-Ziv algorithm (the array is saved using numpy.savez\_compressed in numpy ver. 1.7 in a unix system)\cite{kaspar, numpy}. Figure \ref{methods}D shows the observables corresponding to a network of size $N=200$ using $T=100$ time samples and they were generated using the same parameter values as in Fig. \ref{methods}B. Despite the fact that the parameters where chosen using $C_{LZ}$ computed for a network of size $N=20$, the same parameter values yield qualitatively similar patterns for a network of size $N=200$. Furthermore, their complexity values are consistent with those of the $N=20$ observables (shown in Fig.  \ref{methods}B): note that the relations $C_{LZ}(I)<C_{LZ}(II)$, $C_{LZ}(II)>C_{LZ}(III)$, $C_{LZ}(III)<C_{LZ}(IV)$, $C_{LZ}(IV)>C_{LZ}(V)$ and $C_{LZ}<C_{LZ}(R)$ hold for both the cases $N=20$ and $N=200$ (and the same holds for $C_{LC}$). Finally, while the values of $C_{LZ, LC}$ scale differently with $T$ for different values of the parameters, it was found that different choices of the number of samples $T \in [50, 200]$ lead to qualitatively similar parametric dependences of this quantity. Larger values of $T$ yield values of $C_{LZ, LC}$ that are better able to distinguish among the different patterns. The largest the sample, the easier for the compressor to distinguish an ordered pattern like (I) from a disordered pattern like (IV). However, once this difference is established, further increments of $T$ do not affect the location of the different patterns in parameter space. With these choices, $C_{LZ,LC}$ is sufficient to detect parameter values for which system (\ref{model}) produces qualitatively different behaviors.  

\begin{figure*}[ht!]
\includegraphics[width=180mm]{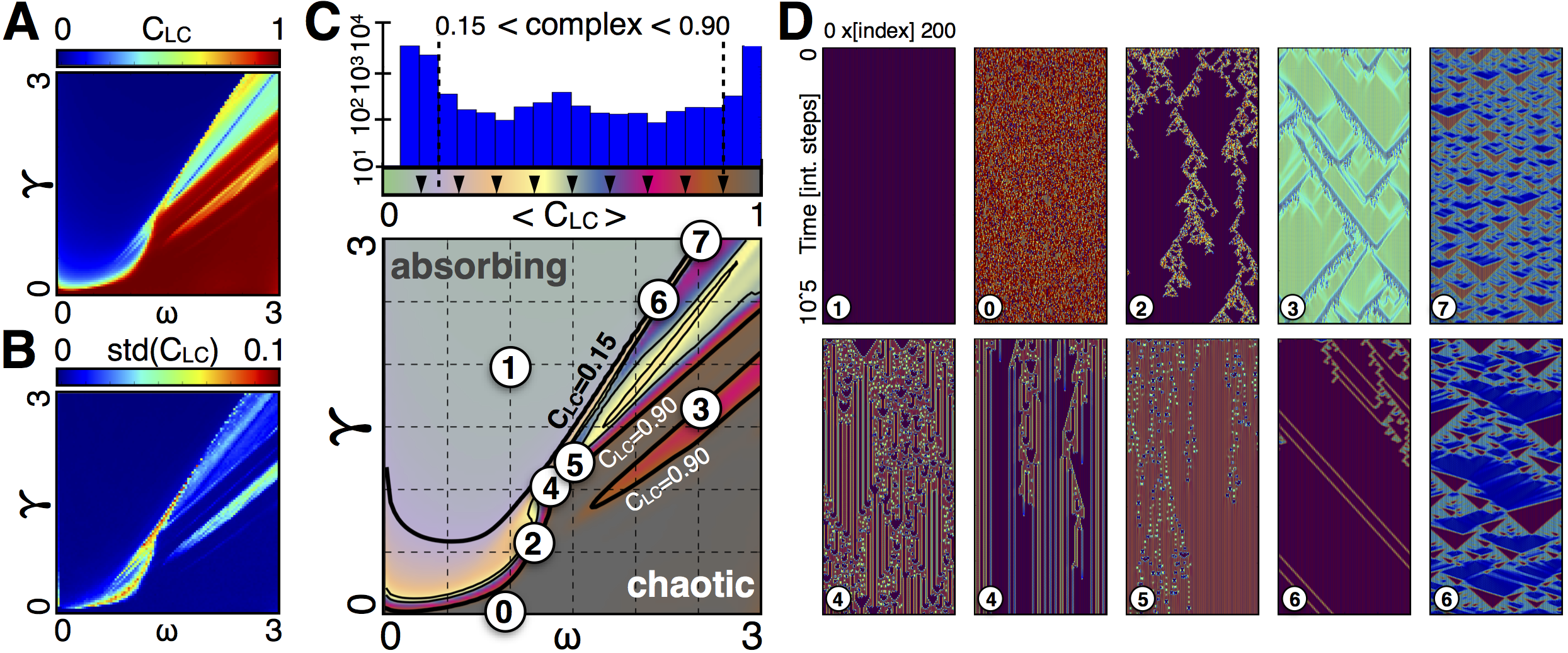}
\caption{\textbf{Phenomenology}. System (\ref{model}) can display a great diversity of spatio-temporal patterns. The complexity $C_{LC}$ of the patterns is measured to show that there are regions where $C_{LC}$ takes similar values and that there are transitions among these regions. \textbf{A} The complexity $C_{LC}$ (in colors) of the solutions of system (\ref{model}) in parameter space averaged over $100$ initial conditions. For each initial condition, $C_{LC}$ was computed in a regularly spaced $101 \times 101$ grid in the inspected domain using $N=200$ units and $T=100$ time samples ($10^4$ int. steps). For most parameter values, $C_{LC}$ is either close to $0$ or close to $1$ and there are regions in parameter space for which $C_{LC}$ takes intermediate values. \textbf{B} Standard deviation of $C_{LC}$ over $100$ initial conditions. For most parameter values, variations in $C_{LC}$ are small ($10^{-2}$). The standard deviation reaches values around $10^{-1}$ in parameter regions in which $C_{LC}$ changes (where $\left|\nabla{C_{LC}}\right|$ is larger).\textbf{C} The histogram shows that the complexity $C_{LC}$ is either smaller than $0.15$ or larger than $0.90$ for most parameter values (y-axis scale is logarithmic, total count is $10^4$). The phase diagram below corresponds to the iso-levels of $C_{LC}$ and highlights $3$ regions: an absorbing region for which all initial conditions decay to an equilibrium ($C_{LC}<0.15$), a chaotic region for which all initial conditions quickly evolve into disordered states ($C_{LC}<0.90$), and smaller regions for which $C_{LC}$ takes intermediate values. Complex behavior can be found in the region between the absorbing and the chaotic regions, and also in regions within the chaotic region. \textbf{D} Spatio-temporal patterns of activity for the points indicated in C. The values of the parameters ($\omega$,$\gamma$,$k=1$) are as follow: \textbf{(1)} $\gamma=1.00,\omega=2.00$, \textbf{(0)} $\gamma=0.00,\omega=1.00$, \textbf{(2)} $\gamma=0.60,\omega=1.20$ , \textbf{(3)} $\gamma=1.65,\omega=2.55$ , \textbf{(7)} $\gamma=3.00,\omega=2.55$ , \textbf{(4l)} $\gamma=1.05,\omega=1.375$, \textbf{(4r)} $\gamma=1.05,\omega=1.350$, \textbf{(5)} $\gamma=1.2,\omega=1.5$, \textbf{(6r)} $\gamma=2.5,\omega=2.2026$, \textbf{(6l)} $\gamma=2.5,\omega=2.2107$. Here, the network ($N=200$) was evolved for $10^5$ integration steps from random initial conditions using $dt=0.01$. The activity $<\boldsymbol{x}> = <\sin(\theta)>$ was averaged over $\Delta=10^3$ time steps and is shown in colors in the range blue $=-0.5$, red $=-0.1$, using the color scale in A and B.}
\label{results1}
\end{figure*}

The results in this subsection are summarized in Figure \ref{results1}. The complexity $C_{LC}$ of system's (\ref{model}) solutions was measured in a $101 \times 101$ regular grid in the inspected range. The size of the system is $N=200$ and the observables $O$ were built using $T=100$ time samples. Figure \ref{results1}A shows the average value of $C_{LC}$ over $100$ random initial conditions. For most parameter values, $C_{LC}$ takes values that are either close to $0$ or to $1$, and there are smaller regions for which $C_{LC}$ takes intermediate values. There are regions in parameter space for which small changes of the parameters lead to qualitatively similar patterns, and there are other regions in which a small change leads to large differences in $C_{LC}$, indicating that the dynamics of the system undergoes qualitative changes. Most notably, $C_{LC}$ displays an abrupt change along a spoon-shaped curve which is the main focus of this article. Figure \ref{results1}B shows the standard deviation of $C_{LC}$ over initial conditions. For values of the parameters where $C_{LC}$ exhibits changes, the standard deviation can be as large as $10^{-1}$, while for most parameter values the deviation is close to $10^{-2}$. For small values of $\gamma < 1$ the complexity $C_{LC}$ exhibits an increase as $\omega \rightarrow 0$ and the standard deviation of $C_{LC}$ also increases. These regimes are not explored in this article and remain the subject of further studies. 

It was found that $C_{LC}$ is an efficient measure to distinguish among the several types of patterns produced by system (\ref{model}). The histogram in the top of Fig. \ref{results1}C shows that the distribution of $C_{LC}$ over the inspected range is dominated by two large peaks near $C_{LC}\approx 0$ and $C_{LC}\approx 1$. There is also a third peak near $C_{LC}\approx 0.5$ that is one order of magnitude smaller and more spread out. In order to distinguish among at least three phases, the interval $0.15 \leq C_{LC} \leq 0.90$ (indicated in the histogram) was chosen so that the two largest peaks of the distribution of $C_{LC}$ are left outside. The diagram in Fig. \ref{results1}C shows $C_{LC}$ in colors and the black solid lines correspond to the iso-values of $C_{LC}$ as approximated by a marching squares algorithm taken at the interval boundaries $C_{LC}=0.15$ and $C_{LC}=0.90$ \cite{marchingsquares}. Different choices of the interval boundaries lead to qualitatively similar diagrams. To aid visualization, the color scale was chosen to highlight regions where $C_{LC}$ takes similar values and the $6 \times 6$ grid was included to provide numerical references. The labels indicate the parameters used in the simulations shown in Figure \ref{results1}D.

System (\ref{model}) displays a multitude of spatio-temporal patterns, some of which are portrayed in Fig. \ref{results1}D. The activity of each unit was averaged over $\Delta = 1000$ integration steps for visualization purposes and is indicated in colors (blue = $-0.5$ to red $ = -0.1$ using the same color scale as in Fig.\ref{results1}A). A first observation is that the activity of the odd units is different from that of the even units, which accounts for the vertical stripes in the patterns. For reference, solution (1) displays an absorbing state in which all initial conditions decay to an equilibrium and solution (0) features a disordered state which corresponds to the non-excitable case $\gamma=0$. A robust feature of this system is the emergence of spatial structures that evolve at a much slower pace than that of the individual units. Complex behavior can be found near the transitions between the complex and chaotic regions as illustrated by solutions (2), (5) and (3), and can also be found near the transitions between the absorbing and complex phases, as illustrated by solutions (6) and (7). While in solutions (2), (5) and (3), only a few units are desynchronized from the background oscillation, in solution (7) all the units are engaged in nested patterns of activity that look self-similar. Near the transitions, a small increment $d\omega = 0.025$ leads to patterns with different behaviors as illustrated by solutions (4l) and (4r). In (4l) large structures propagate on a background of regular oscillations while in (4r) the structures are long-lived but eventually freeze to a non-trivial oscillatory pattern that is spatially inhomogeneous (unlike in (2) and (5)). The network can display traveling pulses as depicted in (6l). After a transient decay the activity settles on a set of pulses that travel with the same speed. A small increment of $d\omega=0.01$ leads to radically different patterns as illustrated by solution (6r). The values of the parameters for all the solutions shown in Fig. \ref{results1}D are indicated in the figure caption.

The solutions of system (\ref{model}) share several features with those of one-dimensional cellular automata (CA). Extensive quantitative studies on 1D CAs have led to a classification scheme into 4 classes \cite{wolfram}. CAs in classes I and II quickly evolve into homogeneous states and stable or oscillating structures. Class III automata evolve in a chaotic manner and Class IV behavior corresponds to the emergence of complex structures that interact in conspicuous ways and may eventually fade \cite{wolfram}. System (\ref{model}) produces solutions that can be classified in a similar way. In Figure \ref{results1}D, solutions (2),(5) and (3) consist of spatially localized structures that interact with each other to produce long transients, resembling type IV behavior. Solutions (6) and (7) exhibit nested patterns of activity that resemble type III behavior. Finally, solutions (4r) and (6l) exhibit an interesting mix of regularity and disorder, and they were found by tuning the parameters of solutions (4l) and (6r) towards the transitions. 

Even in the simplest cases of two-state units and nearest neighbor rules, CAs exhibit remarkable properties at the aggregate level, including the capacity of of supporting universal computations \cite{vonneuman}. A well-known example of such behavior is Conway's \enquote{Game of life}, a two-state 2-dimensional CA whose emergent properties have fascinated researchers in artificial intelligence and computational biology for decades \cite{conwaygardner}. In the \enquote{Game of life}, different initial patterns lead to qualitatively different behaviors: if this CA is started from random initial states the activity will in general decay to quiescence, but if the initial conditions are chosen carefully the system is capable of implementing a Turing machine \cite{rendell}. Not every CA possesses these properties. A study by Langton suggests that CAs that are capable of supporting universal computation (aka. class IV CAs) are typically found in special values of the parameters for which there is a phase transition into chaotic behavior \cite{langton}. To further draw an analogy with Class IV behavior in the system presented here, note that the standard deviation of $C_{LC}$ is largest near the transitions. This indicates that the complexity of the solutions depends more strongly on the initial conditions in those areas. While for most parameter values all initial conditions quickly evolve into qualitatively similar patterns (absorbing or chaotic), for regions of the parameters near the transitions, system (\ref{model}) produces complex patterns whose evolution (and decay times) depends critically on the initial state.  

\subsection{Existence and stability of equilibria}
System (\ref{model}) is a many-body nonlinear problem and despite its simple definition, a full description of its behavior poses a formidable mathematical challenge beyond the scope of this work. However, insight can be gained from an analysis of existence of equilibria. In this subsection I show that system (\ref{model}) undergoes a $N$-dimensional local bifurcation as its control parameters are changed. Furthermore, the location of the bifurcation curve in parameter space provides an explanation for some of the transitions featured in the phase diagram in Fig \ref{results1}C.

\begin{figure}[ht!]
\includegraphics[width=86mm]{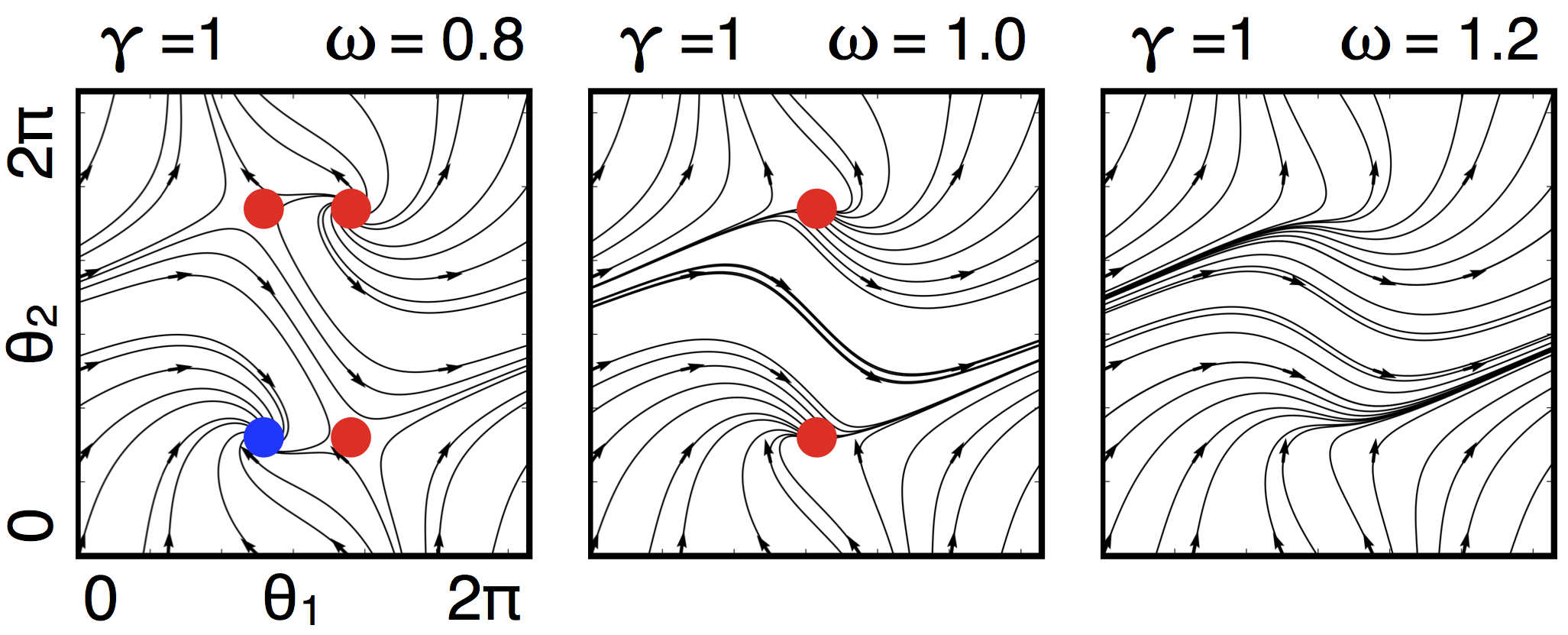}
\caption{\textbf{Phase portrait for $N=2$}. The dynamics of system (\ref{model}) is represented in phase space for the case $N=2$. The arrows indicate the direction of the flow and the filled dots represent the unstable (red) and stable (blue) equilibria. The system can display 4 equilibria that are born out of a pair of saddle node bifurcations on a 2-torus.}
\label{figcase2}
\end{figure}

Analysis of the case $N=2$ is useful to understand larger values of $N$. In the case $N=2$ the phase space of the system is a 2-torus and the equations of motion are given by 
\begin{eqnarray}
  \dot{\theta_1} &=& \omega + \gamma \cos(\theta_1) - k \cos(\theta_2)  \\ \nonumber
  \dot{\theta_2} &=& \omega + \gamma \cos(\theta_2) + k \cos(\theta_1).
\label{case2bis}
\end{eqnarray} 
For $\omega,\gamma,k >0$ the equilibria of eq. (3) are given by 
\begin{eqnarray}
  \cos(\theta^*_1) &=& - \omega \frac{\gamma  + k}{\gamma^2 + k^2} = a_2 \\ \nonumber
  \cos(\theta^*_2) &=&  \omega  \frac{-\gamma + k}{\gamma^2 + k^2} = b_2
\end{eqnarray}
from which the conditions for the existence of equilibria are that $\lvert a_2 \rvert \leq 1$ and $\lvert b_2 \rvert \leq 1$. Using that $\lvert b_2 \rvert < \lvert a_2 \rvert \leq 1$ then system (3) displays equilibria if and only if 
\begin{equation}
  \omega \leq \frac{\gamma^2 + k^2}{\gamma + k}.
\end{equation}
Because the system is defined on a torus, there are in general two solutions for each inversion of the function cosine. The equilibria are given by 
\begin{eqnarray}
  \theta^*_1 &=& \pm \arccos(a_2) \\ \nonumber
  \theta^*_2 &=& \pm \arccos(b_2).
\end{eqnarray}
If $\lvert a_2 \rvert \leq 1$ there are two values for each component of $\theta^*$ totaling $4$ equilibria. If $\lvert a_2 \rvert = 1$ then $\theta^*_1 = \pm \pi$ which corresponds to the same coordinate so in this case there are $2$ equilibria. Because $\lvert b_2 \rvert < \lvert a_2 \rvert$ the situation in which there is only $1$ equilibrium does not occur. The dynamics of system (3) is better visualized in phase space. Figure \ref{figcase2} portrays the three possible scenarios using $\gamma=k=1$ as fixed and considering $\omega$ as the bifurcation parameter. For $\omega<1$ there are $4$ equilibria, one of which is stable (the one that corresponds to the positive branch of the function $\arccos$), while the rest are unstable. Each pair of equilibria is borne out of a saddle-node bifurcation: the structurally unstable steady solutions ($\omega=1$) disintegrate in two pairs of equilibria.

A similar analysis can be performed for $N\geq4$. For this it will be convenient to consider system (\ref{model}) in vector notation. The state of the system can be specified by a $N$-dimensional vector $\boldsymbol{\theta}$ and the dynamics can be expressed as:  
\begin{eqnarray}
\label{modelvector}
\frac{d \boldsymbol{\theta}}{dt}  = \omega + \gamma \boldsymbol{cos}(\boldsymbol{\theta}) + k  \boldsymbol{C} \boldsymbol{cos}(\boldsymbol{\theta}). 
\end{eqnarray}
\begin{figure*}[ht!]
\includegraphics[width=180mm]{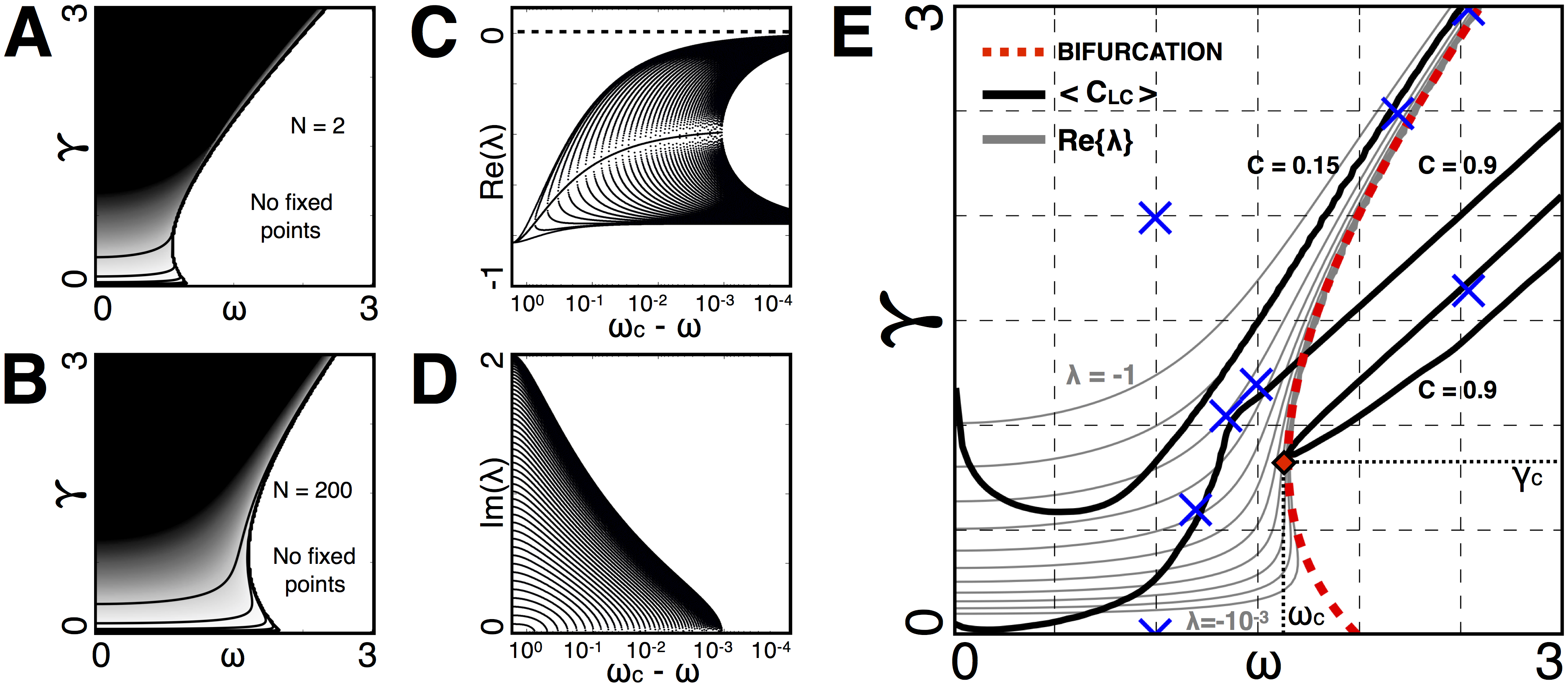}
\caption{\textbf{Existence and stability of equilibria compared to $C_{LC}$.} In terms of existence of equilibria, there are two regions in parameter space: one in which there are no equilibria and another in which there are exponentially many unstable equilibria along with one stable equilibrium. The stability of the stable equilibrium is governed by the real part of the largest eigenvalue $\lambda$ of the Jacobian matrix $J_0$. The real part of $\lambda$ (indicated in gray scale) approaches $0$ as the parameters approach the transition between regions. \textbf{A} Existence and stability of equilibria for $N=2$. \textbf{B} Existence and stability of equilibria for $N=200$. \textbf{C} Linear stability of the stable state for $N=200$. The real part of the eigenvalues of $J_0$ approach $0$ as $\omega$ approaches $\omega_c = 4(\sqrt{2}-1)$ ($\gamma_c = 2(\sqrt{2}-1)$). The horizontal axis corresponds to the distance to the critical value $\omega_c$ in logarithmic scale. For low values of $\omega$ most eigenvectors have the same linear stability. However, as $\omega \rightarrow \omega_c$ the imaginary part of the eigenvalues vanishes yielding different stabilities for different eigenvectors. \textbf{D} Imaginary part of the eigenvalues of the stable state. The imaginary part of the eigenvalues vanishes for $\omega > \omega_c - 10^{-3}$. \textbf{E} Existence of equilibria and stability of the stable state compared to the complexity of the spatio-temporal patterns $C_{LC}$ for $N=200$. The black bold lines indicate the iso-levels of $C_{LC}$ and the dashed bold red line corresponds to the bifurcation curve. The iso-levels of the stability $\lambda$ are plotted in thin gray lines in the range $[-1, 10^{-3}]$. For reference, the blue crosses indicate the parameters used for the simulations in Fig. \ref{results1}D.}
\label{results-fixedpoints}
\end{figure*}  
In this notation the function cosine is applied to each component of vector $\boldsymbol{\theta}$, ie: $\boldsymbol{cos}(\boldsymbol{\theta})_i = \cos(\theta_i)$ ($i \in [1,N]$). The interaction matrix $\boldsymbol{C}$ is anti-symmetric ($C_{ij}=-C_{ji}$) and tridiagonal, 
\begin{equation}
\boldsymbol{C}=\begin{bmatrix}
    0 & -1  & 0  &\hdots   & -1 \\
    1 &  0  & 1 & \ddots   &\vdots       \\
    0 & -1  & \ddots    & \ddots & 0  \\    
    \vdots & \ddots &\ddots & 0 & -1 \\
    1 & \hdotsfor{1}& 0 &1 & 0 \\
\end{bmatrix}.
\end{equation}
Periodic boundary conditions are imposed by setting $C_{1,N} = -1$ ($C_{N,1} = 1$). From eq. (\ref{modelvector}) we see that solving for the equilibria entails solving a linear system of equations. Let $\boldsymbol{\gamma} = \gamma \boldsymbol{I_{N\times N}}$, then 
 \begin{equation}
  -\boldsymbol{\omega} = ( \boldsymbol{\gamma} + k \boldsymbol{C}) \boldsymbol{cos(\theta)}.
\label{eqfixedpoints}
\end{equation}
Assuming that the solutions of eq. (\ref{eqfixedpoints}) have the form $\cos(\theta_b) = \cos(\theta_{2i}) = b $ and $\cos(\theta_a) = \cos(\theta_{2i+1}) = a$ and inserting these expressions in eq. (\ref{eqfixedpoints}) yields the solution
\begin{eqnarray}
a &=& -\omega \frac{\gamma + 2 k}{\gamma^2 + 4 k^2} \\ \nonumber
b &=& -\frac{\omega + 2 k a }{\gamma}. 
\label{fpsols}
\end{eqnarray}
Therefore, the conditions for the existence of equilibria are that $\lvert a \rvert \leq 1$ and $\lvert b \rvert \leq 1$. It is convenient to define $g(\gamma, k) = \frac{b}{a}$ which simplifies to $g(\gamma, k) = \frac{\gamma^2 - 2k \gamma}{\gamma^2 + 2k \gamma}$. Using that $\displaystyle{\lim_{\gamma \to \infty}} g(\gamma,k) = 1$,$\displaystyle{\lim_{\gamma \to 0}} g(\gamma,k) = -1$ and $\frac{dg}{d\gamma}>0$ $\forall k>0$ then $\lvert g \rvert <1$ implies that 
$\lvert b \rvert < \lvert a\rvert$. This means that we need only to consider the condition $\lvert a \rvert \leq 1$. Thus we see from eq. (10) that equilibria exist if and only if,
\begin{equation}
\omega \leq \frac{\gamma^2 + 4 k^2}{\gamma + 2 k}.
\label{condition}
\end{equation}
If condition (\ref{condition}) is met, then the equilibria are given by 
\begin{eqnarray}
\theta^*_{2j-1} &=& \pm  \arccos(a) \\ \nonumber
\theta^*_{2j} &=& \pm    \arccos(b)
\label{fixedpoints}
\end{eqnarray}
where $j = \{1 ,  . .  , \frac{N}{2}\}$. From this equation we see that if $\lvert a \rvert < 1$ there are $2^{\frac{N}{2}}$ possible values for the odd components of $\theta^*$ and $2^{\frac{N}{2}}$ values for the even components, yielding $2^N$ equilibria. If $\lvert a \rvert = 1$ then $\theta_{2i-1} = \pm \pi$ which indicates that all the odd components are identical, and therefore there are $2^{\frac{N}{2}}$ equilibria. As in the case $N=2$, each of these equilibria undergo a saddle-node bifurcation and give rise to a pair of equilibria. Thus, an exponentially large number of equilibria are created for parameter values that satisfy the equality in eq. (\ref{condition}). The bifurcation curve is given by
\begin{equation}
\omega = \frac{\gamma^2 + 4 k^2}{\gamma + 2 k}.
\label{transition}
\end{equation}
 
The linear stability of the equilibria is determined by the derivative of eq. (\ref{modelvector}). The equilibria can be indexed by taking the positive branch of function $\arccos$ as a reference, and defining a vector $\boldsymbol{m}$ such that if $\theta^*_i<0$ then $m_i =1$ and $m_i = 0$ if $\theta^*_i \geq 0$. Then each equilibria is mapped to an integer $M$ with binary representation $\boldsymbol{m}$ defined by $M = \sum^{N-1}_{k=0} m_k 2^k$. There are two states that have the symmetries of regular polygon with $\frac{N}{2}$ sides  (dihedral symmetry $D_{N/2}$): these correspond to taking all components of $\theta^*$ to be positive ($M=0$) and to taking all components of $\theta^*$ to be negative ($M=2^N$). Evaluation of the derivative of eq. (\ref{modelvector}) yields a family of matrices, 
\begin{equation}
  \boldsymbol{J}_M = -(\boldsymbol{\gamma} + k \boldsymbol{C}) \boldsymbol{\sin(\theta_M^*)}, 
\end{equation}
where $\boldsymbol{\sin(\theta_M^*)}_i = \sin((-1)^{m_i}\theta_i^*)$. Here $\boldsymbol{J}_M$ is a family of tridiagonal matrices and a thorough analysis of their spectrum is beyond the scope of this work (see ref. \cite{jacobi} for a mathematical discussion of this problem). In this work the spectrum of $J_M$ was explored numerically. The state $M=0$ is stable for all values of $N \in [2,1000]$. This was verified numerically and a proof of this statement (if true) for any (even) $N$ could not yet be found. It was also found numerically that the state $M=0$ is the only stable state and the state $M=2^N$ has no stable directions. The uniqueness of the stable state (if true) could also not be demonstrated. Because the number of equilibria increases exponentially with $N$, the uniqueness of the stable state can not be verified numerically by exhaustion for large $N$. However, it can be established numerically that the probability that there is another stable state is smaller than $10^{-6}$ (for $N=200$). Whether the stable state is unique or not is not central to this article but a rigorous proof of this statement (if true) is highly desirable for future investigations. The state $M=0$ (and $M=2^N$) is unique because of its symmetries so this criteria is adopted to focus on the stability properties of this particular state. 

Quantification of the complexity of system's (\ref{model}) solutions reveals an organization in parameter space with at least three different regions. In order to associate the transitions among these regions with the dynamical properties of the system, the stability of the stable state $M=0$ was explored numerically. This is shown in Figs. \ref{results-fixedpoints}A and \ref{results-fixedpoints}B for $N=2$ and $N=200$. The real part of the largest eigenvalue $\lambda_m$ of $\boldsymbol{J_0}$ (ie: $\lambda_m =\max_{\lambda} \{Re(\lambda)\}$ with $\lambda$ an eigenvalue of $\boldsymbol{J_0}$) is shown in gray scale (black $=-3$ white $=0$). In terms of existence of equilibria, the system features two regions: one where there are no equilibria and another with an exponentially large number of unstable equilibria and one stable equilibria. The transition between these regions is given by equation (\ref{transition}) for all (even) $N \geq 4$. Figures \ref{results-fixedpoints}C and \ref{results-fixedpoints}D show the real (and imaginary) parts of the eigenvalues of $J_0$, as $\omega$ (in logarithmic scale) approaches a point in the transition curve ($\omega_c = 2 \gamma_c$ with $\gamma_c= 2 (\sqrt(2)-1 )$). For values of $\omega$ away from the transition, the real part of the eigenvalues take one of a small set of possible values while the imaginary part takes a larger set of values. The situation is reverted as $\omega \rightarrow \omega_c$ and the imaginary part of some eigenvalues vanishes, yielding a pair of eigenvalues with different real parts. The imaginary part of all eigenvalues vanishes between $\omega_c - 10^{-3}$ and $\omega_c - 10^{-4}$.

Figure \ref{results-fixedpoints}E compares the existence and stability of the stable equilibrium to the complexity of the solutions. The iso-levels of $C_{LC}$, indicated by thick black lines, are compared with the bifurcation curve given by eq. \ref{transition}, plotted with a thick dashed red line. The iso-levels of $\lambda_m$ are plotted with thin gray lines. Note that as $\omega \rightarrow 3$, both the bifurcation curves and the curve $\lambda_m=-1$ approximate the curve $C_{LC}=0.15$ which denotes the transition between the absorbing and complex phases as defined in Fig. \ref{results1}C. It is convenient here to consider the limiting cases given by $k=0$ and $\omega=\gamma \rightarrow \infty$. In the absence of interactions ($k=0$), the bifurcation curve given by eq. (\ref{transition}) becomes $\omega_b=\gamma_b$, and parameter space is again split in two regions: in one region ($\omega_b<\gamma_b$) there is one equilibria that will yield observables with low complexity, while in the other region ($\omega_b>\gamma_b$) there are uncorrelated, identical oscillations, that yield larger values of the complexity than a decay to the absorbing state. Therefore, for the case $k=0$ the bifurcation curve would be sufficient to (trivially) explain the location of different phases. In the case that the inspected range is very large ($\omega=\gamma \rightarrow \infty$), we see from taking the limit $\gamma \rightarrow \infty$ in eq. (\ref{transition}) that the bifurcation curve converges asymptotically again to $\omega_b=\gamma_b$: the interactions will not play a role in the limit that the natural frequencies and excitations are infinitely large, unless their magnitudes are comparable (ie. near $\omega_b=\gamma_b$). Thus, the results presented here suggest that the phase diagram in Fig. \ref{results1} can be partly explained by the bifurcation curve given by eq. (\ref{transition}), and indicate that such agreement is more explicit as $k \rightarrow 0$. 

The equilibria of system (\ref{model}) are born arbitrarily close to each other in pairs from multiple saddle-node bifurcations: therefore, the smooth field induced by eq. (\ref{model}) becomes arbitrarily small along the direction that joins each pair of equilibria. On the other side of the bifurcation curve (\ref{transition}) the field also becomes arbitrarily small in neighborhoods of state space where equilibria are born (the ghosts of the saddle-node bifurcations). If the state of the system visits such neighborhoods, it will spend more time there than in other regions, again providing a means for the system to produce multiple time-scales. The results in this article show that near the bifurcation curve (\ref{transition}) complex behavior is associated with large numbers of equilibria, which, upon changes of the parameters, undergo a local (saddle-node) bifurcation. Complex behavior occurs near the bifurcation curve (\ref{transition}) and on one side of it, but not in the absorbing region. This suggests that the presence of equilibria alone is not sufficient to enable complex behavior and that the slower dynamics near the equilibria, when the parameters are close to the bifurcation, may play a role in the global dynamics. Finally, as shown in the phase diagram in Fig. \ref{results1}C, the landscape of $C_{LC}$ presents regions of complex behavior away from the transition as well as other features that cannot be explained by the arguments presented here and remain a subject of future investigation.

\subsection{Extension to two-dimensional lattices}
For clarity, the analysis in this article was restricted to a one-dimensional lattice or chain. This subsection shows that the model can be generalized to spatially extended lattices of higher dimensions and that the analysis can be extended to those cases. In particular, this subsection focuses on an extension of system (\ref{model}) to the case of a two-dimensional lattice and provides numerical evidence that system (\ref{model}) is a useful case study for pattern formation.

An extension of the model to a two-dimensional lattice is given by, 
\begin{eqnarray}
\dot{\theta_{i,j}}&=&\omega + \gamma \cos(\theta_{i,j}) + k (-1)^{i+j} F_{ij}\\ \nonumber
F_{ij} &=&\{\cos(\theta_{i-1,j}) + \cos(\theta_{i+1,j}) \\ \nonumber
       &+&\cos(\theta_{i, j-1}) +\cos(\theta_{i, j+1})\}. 
\label{model2d}
\end{eqnarray}
In order to balance the strength of the interactions in every local patch, the units are arranged in an alternating \enquote{chessboard} configuration via the term $(-1)^{i+j}$ and periodic boundary conditions are set by $\theta_{i,1} = \theta_{i,N}$ and $\theta_{1,j} = \theta_{N,j}$. As in the one-dimensional case, all units are either purely excitatory or purely inhibitory, and also, all units receive only excitatory or inhibitory input. As before, the equilibria of system (15) are given by $\cos(\theta^*_{i+j}) = a$ if $(i+j)$ is even and $\cos(\theta^*_{i+j}) = b$ if $(i+j)$ is odd. Therefore, by replacing $k$ with $\frac{k}{2}$, the analysis in the previous section applies to the two-dimensional case (and to the N-dimensional case by the same arguments).

\begin{figure}[ht!]
\includegraphics[width=86mm]{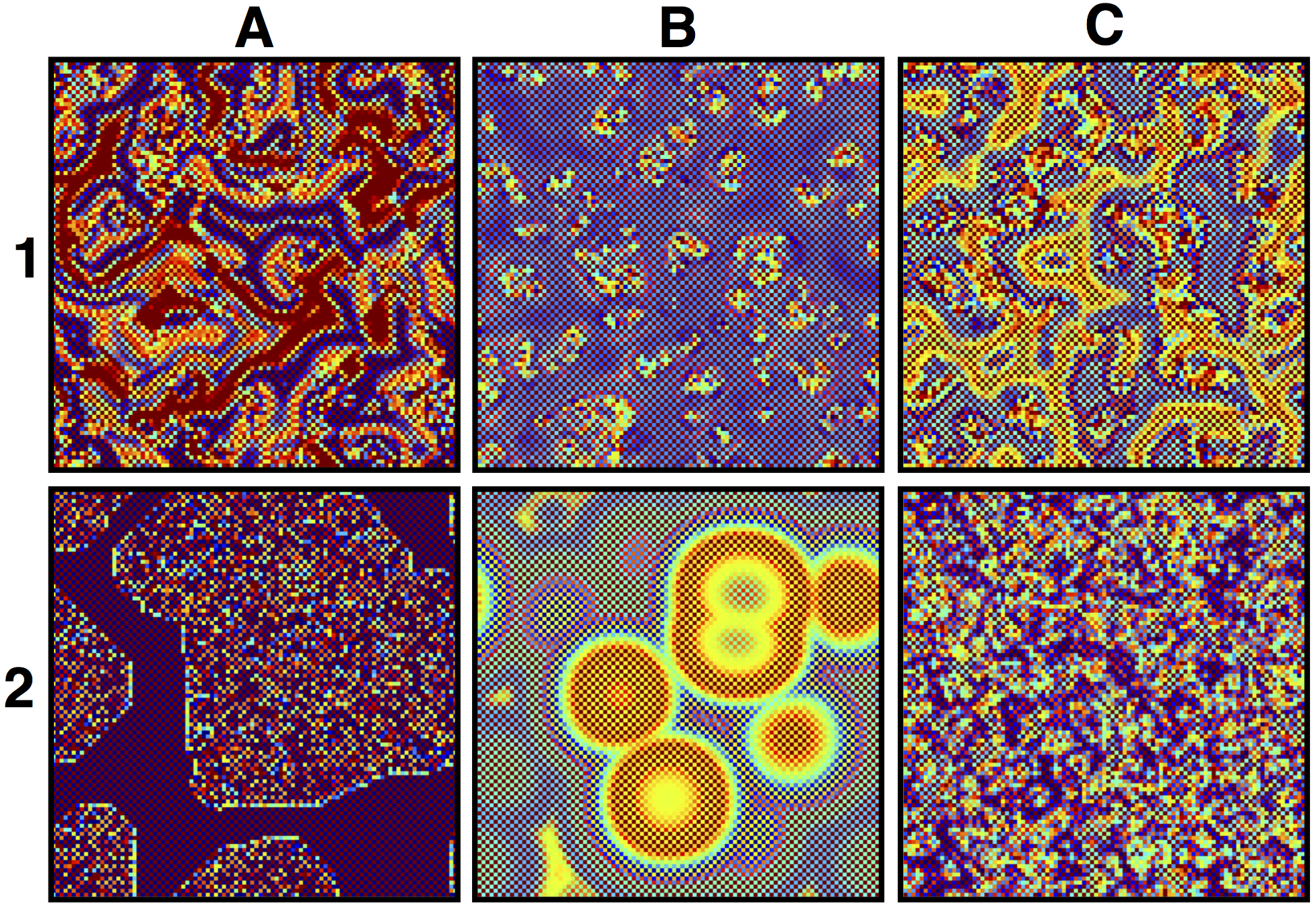}
\caption{\textbf{Pattern formation in a network of excitable phase oscillators.} The units are placed on a $100 \times 100$ two-dimensional lattice and the interactions are local. The panels show the solutions of system (15) at time $t_0= 1.6 \times 10^5$ (int. steps) for different values of parameters $(\omega, \gamma,k)$. At each site of the lattice, the activity of the corresponding oscillator is indicated in colors (the color scale is the same in all panels). The parameters for the simulation are: ($(k=0.5)$ in all panels), (1A) $(\omega=2, \gamma=0.77)$, (1B) $(\omega=1.8, \gamma=0.9)$, (1C) $(\omega=1.5, \gamma=0.7)$, (2A) $(\omega=1.1, \gamma=0.8)$, (2B) $(\omega=1.6, \gamma=0.8)$, (2C) $(\omega=2, \gamma=1)$. (Multimedia view)}
\label{2dpatterns}
\end{figure} 

As in the one-dimensional chain, for different values of the parameters, system (15) can display a multitude of different spatio-temporal patterns. This is illustrated in Figure \ref{2dpatterns}. In all panels, the network is defined on a $100 \times 100$ lattice ($N=10^4$) and was evolved for $t=1.6 \times 10^5$ integration steps starting from random initial conditions (the parameter values are shown in the Figure's caption). In solutions (A1), (C1) and (C2) most units are engaged in spatio-temporal patterns that feature different spatial scales. This also occurs in solution (B2), but in this case the patterns are more regular. In solution (B1), most units are engaged in a regular oscillatory background and fewer units desynchronize from the background activity to give rise to structures that propagate in timescales much slower than the natural period of the oscillators. Finally, solution (A2) presents a striking mix between order and disorder. In two regions the activity of the oscillators is chaotic and there are no evident regularities. Outside these region the units remain quiescent. The boundary that separates these regions changes its shape very slowly when compared to the much faster activity that takes place inside the disordered regions.

Despite its simple definition, system (15) is capable of generating a wide diversity of spatio-temporal patterns, and, because it can be placed in a two-dimensional surface, for large $N$ it may be useful as a discrete approximation to describe the evolution of a continuous quantity on that surface. Also because the interactions are local, system (15) may be useful as a case study to address pattern formation in spatially extended physical systems with diffusive coupling, such as the oscillatory chemical reactions that produce propagating reaction-diffusion fronts\cite{turing,epsteinshowalter}.

\section{Conclusions}
The purpose of this article is to explore sufficient conditions under which an assembly of oscillatory units can display complex behavior. The system can ---for open ranges of the parameter values--- display dynamical patterns of activity which spread from traveling pulses to complex emergent structures. The model features at least three phases: an absorbing phase in which all initial conditions decay to equilibrium, a chaotic phase in which all initial conditions quickly evolve into disordered states, and an intermediate phase in which the solutions display complex spatio-temporal patterns. Upon changes of its parameters, system (\ref{model}) undergoes a high-dimensional bifurcation where an exponentially large number of equilibria are created locally. The location of the bifurcation curve partly explains the location of the transition between the absorbing and complex phases. Therefore, in this model, complex behavior can be associated explicitly with the presence of large numbers of equilibria with marginal stability. 

The model presented here suggests interesting theoretical implications. Because the units of the system are phase oscillators the results may be applicable to other families of oscillators for which a reduction to phase equations is possible. Complex dynamics should occur robustly in systems of weakly interacting oscillators that meet two conditions: first, in the absence of interactions ($k=0$), the units should undergo a SNILC bifurcation as a control parameter is changed, and second, the interactions poise the network to a critical state in which long-range correlations are possible. In system (\ref{model}), the first condition is met by introducing the excitation term and the second condition is achieved by the assumption that the strength of the interactions between neighboring units is identical and that it is balanced locally. While there may be other choices for the interactions which could tune the system to a critical state, balancing excitation and inhibition locally is arguably a simple way to achieve this. This choice is motivated both by mathematical considerations and by the growing evidence that many biological systems do operate in a critical regime \cite{magnascoprl,bialek}. In the brain, for instance, it is well established that nervous processes rely upon local interactions of excitatory and inhibitory cells \cite{wilsoncowan}. Brain cells are different from other cells in many regards, but perhaps most notably because of their intrinsic dynamical properties, which include the presence of several types of excitability \cite{hh,izi200}. The results presented here suggest that such differentiated dynamics together with a local balance of excitation and inhibition are sufficient to support complex dynamical states, which may in turn be a desirable feature for a neural network. Interestingly, the results in this article are largely independent of the size of the system. This may render the assumptions in the model of potential relevance to the study of networks composed of a few hundreds of units, such as the neuronal networks that control the behavior of nematodes \cite{mei2015}.    

For wide ranges of the parameter values, the dynamics of the system resembles that of cellular automata. The solutions of system (\ref{model}) can be classified as absorbing, chaotic and complex. At the transitions among these regions, the system can display complicated spatial structures that evolve and interact on time scales much larger than that of the natural period of the oscillators. While a computer simulating a CA can be regarded as being described by a (complicated) deterministic, time-continuous equation, system (\ref{model}) may point to a connection between the dynamics of CA and a much simpler time-continuos system. Because of its simplicity, system (\ref{model}) is interesting from a purely mathematical perspective, but it also has potentially important engineering applications. The results in this article show that complex spatio-temporal patterns are achievable in a simple differential equation: this suggests that an electronic implementation of system (\ref{model}) may be possible, and also scalable because the units and the interactions are identical. The dynamics of the physical units utilized in present-day computers ultimately rely on some sort of bi-stability between the states that correspond to the logical on and off states \cite{landauer}. Analogously, in this system the units can be entrained into one of many stable periodic patterns suggesting the capacity to support many-state discrete functions. Future investigations will explore the possibility that system (\ref{model}) could be controlled to perform useful computations. 

\section{Acknowledgments}
Leandro M. Alonso's research was partly supported by funds from a Leon Levy Fellowship at The Rockefeller University. The author acknowledges the valuable reviews of two anonymous referees.

\end{document}